\documentstyle[pre,aps,multicol,epsf]{revtex} 
\begin{document}
\title{Landau Damping and Coherent Structures in Narrow-Banded
1+1 Deep Water Gravity Waves}
\author{ Miguel Onorato, Alfred Osborne, Marina Serio}
\address{Dipartimento di Fisica Generale, Universit\`{a} di Torino, Via
Pietro Giuria 1, 10125 Torino, Italy}
\author{Renato Fedele}
\address{\it Dipartimento di Scienze Fisiche, Universit\`{a} Federico II
di Napoli and INFN Sezione di Napoli\\ Complesso
Universitario di M.S. Angelo, Via Cintia, I-80126 Napoli,
Italy}
\date{\today}
\maketitle
\begin{abstract}
We study the nonlinear energy transfer around the peak of the spectrum of
surface gravity waves by taking into account nonhomogeneous effects. In the
narrow-banded approximation the kinetic equation resulting from a
nonhomogeneous wave field is a Vlasov-Poisson type equation which includes
at the same time the  random version of the Benjamin-Feir instability and
the Landau damping phenomenon. We analytically derive
the values of the Phillips' constant $\alpha$ and the
enhancement factor $\gamma$ for which the narrow-banded approximation
of the JONSWAP spectrum is
unstable. By performing numerical simulations of the
nonlinear Schr\"{o}dinger equation we check the validity of the prediction of
the related kinetic equation. We find that the effect of Landau damping
is to suppress the formation of coherent structures. The problem of
predicting freak waves is briefly discussed.
\end{abstract}
%\pacs{PACS numbers\,: 47.35.+i, 47.20.-k, 92.10.Hm}]%
%
%03.Kf, 03.40.Gc, 47.20.Ky, 47.35.+i}] %
%{PACS Numbers:47.27.Nz}
%
\begin{multicols}{2} 
In many different fields of nonlinear physics, local
nonlinear effects such as the modulational instability
(MI) have played a very important role in spectral energy
transfer processes (see for example \cite{02,05,07}). For
ocean gravity waves, the same instability (now commonly
known as the Benjamin-Feir instability) has been
discovered independently by Benjamin and Feir \cite{BF}
and by Zakharov \cite{ZAK} in the sixties. The
instability predicts that in deep water a monochromatic
wave is unstable under suitable small perturbations. In
this framework it has been established that the MI can be
responsible for the formation of freak waves
\cite{TRU,TDF,ONO}. The aim of this Letter is to discuss
a statistical approach that allows one to predict the
appearance of the modulational instability in a
continuous spectrum. The theory is applied directly to
ocean waves but it could as well be applied to many other
nonlinear wave systems.

In order to approach statistically the nonlinear energy transfer processes
involved in the MI, one is interested in finding a suitable kinetic
description. As far as ocean waves are concerned, today the most common
models for wave forecasting (WAM, SWAM and newer generation models, see \cite
{KOM}) are based on the nonlinear energy transfer process that is ruled by
the {\it kinetic wave equation} that has been derived independently by K.
Hasselmann \cite{HAS} and by V. Zakharov \cite{ZAK}. The theory predicts
that the energy is transferred in an irreversible manner under the resonant
4-wave interaction.  Besides the quasi-gaussian approximation \cite
{BEN}, one of the major hypothesis required to derive this kinetic equation
is the homogeneity of the surface, i.e., $<A(k)A^*(k^{\prime})>=n(k)%
\delta(k-k^{\prime})$, where $A$ is a complex wave amplitude describing the
evolution of the wave train, $k$ and $k^{\prime}$ are wave numbers and $n(k)$
is the spectral density function. Unfortunately, the Hasselmann-Zakharov
kinetic theory is not able to predict the Benjamin-Feir instability. This
basically means that the kinetic equation is unable to predict local
nonlinear effects such as freak waves. Indeed the MI is the result of an
interaction of waves that are phase-locked: the carrier wave is phase-locked
with the side-bands and therefore the mechanism cannot be predicted by the
assumption that the Fourier components are delta-correlated. Thus, if the
hypothesis of homogeneity of the system is relaxed, an improved kinetic
equation can be derived which is able to show a random version of the
Benjamin-Feir instability. Actually, for surface gravity waves, this
improvement is contained in the pioneering work by Alber \cite{ALB},
followed by the works of Crawford et al. \cite{CRO} and Janssen \cite
{JAN1,JAN2}. The prediction of the theory developed in \cite{ALB,CRO,JAN1}
has never been verified numerically or experimentally. Unfortunately these
works, written more than 15 years ago, did not receive the deserved
attention by the wave forecast/hindcast community and the ideas developed
remained basically unconsidered with the only exception of the recent work
by M. Stiassnie \cite{STIA}. More recently, but independently of the works
\cite{ALB,CRO,JAN1}, a similar approach has been developed for the
large-amplitude electromagnetic wave-envelope propagation in nonlinear media
\cite{012}, for the quantum-like description of the longitudinal
charged-particle beam dynamics in high-energy accelerating machines in the
presence of an arbitrary coupling impedance \cite{013} and for the resonant
interaction between an instantaneously-produced disturbance and a partially
incoherent Langmuir wave \cite{014}.

In this paper, we outline the approach formulated in
\cite{ALB,CRO,JAN1} and discuss the modulational
instability for random wave spectra.  In particular we
identify the values of the parameters of the JONSWAP
spectrum for which the spectrum itself is unstable.
Moreover, using numerical simulations we associate the MI
with the presence of coherent structures in physical
space, pointing out the importance of the role played by
the phenomenon of Landau damping.

We start by considering the simplest and prototypical weakly nonlinear
equation for water waves: the Nonlinear Schroedinger equation (NLSE).  This
choice is dictated by the following motivations. First of all, we want to
avoid dealing with the dynamics of the four-wave resonant interactions
because otherwise we would not be able to discern whether the effects we are
describing are due to inhomogeneity or are already contained in the
Hasselmann-Zakharov theory. We point out that, while the two dimensional
NLSE (and higher order equations (see \cite{TKDV})) contains the four-wave
resonant interaction mechanism \cite{LONG}, the one-dimensional NLSE does
not include this sort of energy transfer, because of the shape of the linear
dispersion relation (see for instance \cite{MMT}). Second, while the NLSE
should not be appropriate for describing the dynamics in the tail of the
spectrum or in the inertial range, it should describe with satisfactory
accuracy the short time scale behavior around the peak. Formally the NLSE
can be derived from the Zakharov equation \cite{ZAK} under the narrow band
approximation; in dimensional form the equation reads: %%
\begin{equation}
{\ \frac {\partial A} {\partial t}+ i \mu \frac {\partial^2 A} {\partial x^2}%
+ i \nu |A|^2 A=0, }  \label{nls}
\end{equation}
where in deep water $\mu=\omega_0/8k_0^2$ and $\nu=\omega_0k_0^2/2$, with $%
\omega_0$ the carrier angular frequency and $k_0$ the respective wave
number. 

Eq. (\ref{nls}) is our starting point for deriving the required kinetic
equation. Following Alber \cite{ALB}, the Wigner-Moyal transform \cite{WIG}
can be applied directly to the NLSE. This transform allows one to give a
representation of a function $A(x)$ both in configuration space, $x$, and in
phase space or wave number, $k$, namely
\begin{equation}
{\ n(x,k)=\frac{1} {2\pi} \int <A^*(x+y/2) A(x-y/2)> e^{-iky} dy .}
\label{wigner}
\end{equation}
Alternatively, one may follow the approach presented both in \cite{CRO} and
in \cite{RUB} consisting in writing the NLSE in Fourier space and then
writing an evolution equation for the correlator $n(k,k^{%
\prime})=<A(k)A^*(k^{\prime})>$. Both approaches need a closure in order to
reduce the fourth-order correlator into the sum of the product of the
two-point correlation functions. Using any of the methods outlined the
resulting kinetic equation is the following von Neumann-Weyl-like equation
\begin{eqnarray}
&& \frac {\partial n(x,k,t)} {\partial t}+ 2 \mu k \frac {\partial n(x,k,t)%
} {\partial x} +4\nu \sum^{\infty}_{m=0} \frac {(-1)^m} {(2m+1)! 2^{2m+1}} {%
{\times}}  \nonumber \\
&& \frac {\partial^{2m+1} <|A(x,t)|^2>} {\partial x^{2m+1}} \frac {%
\partial^{2m+1} n(x,k,t)} {\partial k^{2m+1}}=0,  \label{transport_1}
\end{eqnarray}
with
\begin{equation}
{\ <|A(x,t)|^2>=\int n(x,k,t) dk .}  \label{transport_2}
\end{equation}
If the limit for small $k$ is taken the equation resembles the
Vlasov-Poisson equation in plasma physics that is well known to describe the
Landau damping phenomenon. The standard way to proceed consists
in letting the distribution function $n(x,k,t)$ and the field $A(x,t)$ be
expressed in terms of an equilibrium value plus a small perturbation and
study the dispersion relation of the linearized equation for the
perturbation. After standard algebra the following dispersion relation is
obtained (see also \cite{012,013,014}):
\begin{equation}
{1 + \frac{\nu} {\mu} \int \frac { n_0(k+K/2)-n_0(k-K/2)} { K
(k-\Omega/(2\mu K) )} dk =0,}  \label{disp_rel5}
\end{equation}
where $n_0$ is the homogeneous envelope spectrum.

\begin{figure} 
\epsfxsize=8.5cm 
\epsfbox{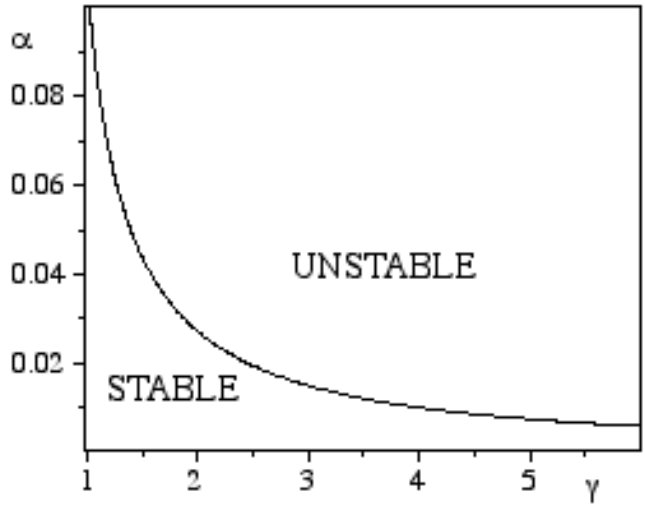}
\caption{Instability diagram in the $\alpha-\gamma$ plane.} \label{fig inst1}
\end{figure}

According to experimental studies, the surface wave spectrum for gravity
waves is given by the JONSWAP spectrum:
\begin{equation}
{P(k)=\frac{\alpha} {2k^3}
e^{-\frac{3} {2} [\frac {k_0} {k}]^2}
\gamma^{exp[-\frac{(\sqrt(k)-\sqrt(k_0))^2} {2\delta^2 k_0}] } ,}
\label{JONSWAP}
\end{equation}
with $\alpha$, $\gamma$ and $\delta$ constants ($\delta$ is usually set to
0.07, while $\alpha$ and $\gamma$ depend on the state of the ocean). In
order to be consistent with the NLSE, as is done in \cite{JAN2}, we
Taylor-expand the spectrum around its peak and obtain the following
Lorenzian spectrum:
\begin{equation}
{P(k)=\frac{H_s^2} { 16 \pi} \frac{p} {p^2+(k-k_0)^2} ,}  \label{lorenzian}
\end{equation}
where

\begin{equation}
{\ p=\sqrt{\frac {8 k_0^2 \delta^2} {24 \delta^2+Log(\gamma)}} \; {\rm and}
\; H_s= 4 \sqrt{\pi \frac{ \alpha \gamma p} { 2 E^{3/2} k_0^{3}}} .}
\label{band}
\end{equation}
$H_s$ is the significant wave height calculated as four times the standard
deviation of the wave field and $p$ corresponds exactly to the half-width at
half-maximum of the spectrum. It can be shown \cite{CRO} that for a
symmetric spectrum $P(k)$ of the surface elevation, the spectrum for the
complex envelope is $n_0(k)=4 P(k-k_0)$, therefore a factor of four must be
taken into account. Substituting in (\ref{disp_rel5}) we obtain the
following dispersion relation:
\begin{equation}
{\Omega=K(\sqrt{K^2\mu^2-H_s^2\nu\mu}-2 i \mu p). }  \label{disp_rel}
\end{equation}
If $K^2<H_s^2\nu/\mu$, the first term on the right hand side is responsible
for the MI (note that in the limit as $p\rightarrow 0$, the dispersion
relation (\ref{disp_rel}) gives the Benjamin-Feir instability). The last
term on the right hand side is responsible for the Landau damping phenomenon
\cite{018}. Therefore there is a competition between exponential growth and
damping of the perturbation that depends on the parameters $\alpha$ and $%
\gamma$ of the Lorenzian (or JONSWAP) spectrum. If $p>\sqrt{-K^2 \mu/4+H_s^2
\nu /4}$ the damping dominates the MI, the opposite will occur if $p<\sqrt{%
-K^2 \mu/4+H_s^2 \nu /4}$. In Fig. \ref{fig inst1} we show the instability
diagram in the $\alpha-\gamma$ plane. In general spectra with higher values
of $\alpha$ and $\gamma$ are more likely to show the MI. This results gives
support to the results in \cite{ONO} where it has been found that higher
values of $\alpha$ and $\gamma$ increases the probability for the formation
of freak waves.
%%%%%%%%%%%%%%%%%%%
%%%%%%%%%%%%%%%%%%%
Before comparing numerical simulations with the prediction of the
Wigner-Moyal-like kinetic equation, we would like to point out that our
analysis has been carried out linearizing around an equilibrium state of the
system. In fact, the dispersion relation (\ref{disp_rel}) has been derived
for small amplitude perturbations. However, in the natural long-time
evolution of a nonlinear wave train, perturbations are in general not small.
Consequently, perturbations in the simulations should evolve according to
the governing equations. To this end, numerical experiments have been
carried out allowing the nonhomogeneities to develop by themselves according
to the nonlinear dynamics of the NLSE.

Numerical simulations of Eq. (\ref{nls}) have been computed using a standard
pseudo-spectral Fourier method. Initial conditions for the free surface
elevation $\zeta (x,0)$ have been constructed as a random process \cite
{OSB_EF}, i.e. a linear superposition of Fourier modes (\ref{JONSWAP}) with
random phases. The Hilbert transform is used in order to convert the free
surface $\zeta$ to the complex envelope variable $A$ of the NLSE. The
dominant wave number for the numerical simulation was selected to be $k_0=0.1
$ $m^{-1}$. This last choice is not restrictive: the parameters that rule
the dynamics in the spectrum are its width and the steepness which, once $k_0
$ is fixed, are univocally determined by $\alpha$ and $\gamma$.

\begin{figure} 
\epsfxsize=8.5cm 
\epsfbox{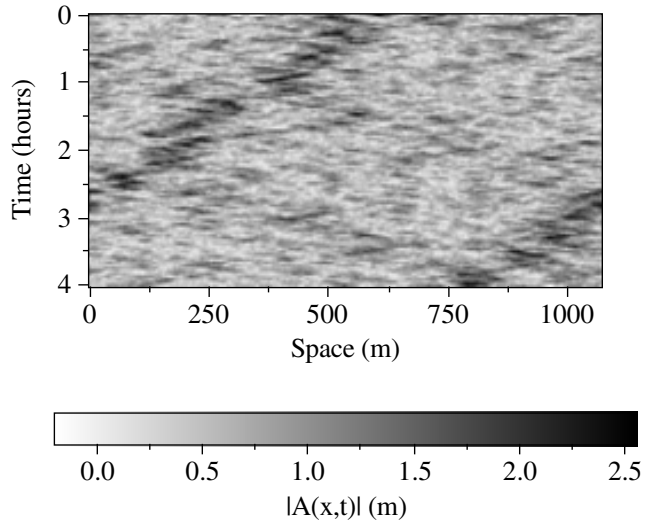}
\caption{$|A(x,t)|$ from numerical simulation of the NLSE. The initial condition
is characterized by a Lorentian spectrum with $\gamma=3$, $\alpha=0.02$.
A coherent structure is evident in the $x-t$ plane.} \label{fig figura1}
\end{figure}

We start the discussion on the numerical results by showing the evolution of
$|A(x,t)|$ in the $x-t$ plane for a case in which $\gamma=3$ and $\alpha=0.02
$ ($p/k_0\simeq0.25$ and $H_s k_0/2 \simeq 0.14$ ), Fig. \ref{fig figura1}.
According to Fig. \ref{fig inst1} the spectrum should be unstable. How is
this instability manifested in configuration space? From Fig. \ref{fig
figura1} we note the presence of a ``coherent structure'', i.e. a structure
(oblique darker zones in the $x-t$ plane) that persists in the presence of
nonlinear interactions and maintains statistically its shape and velocity
during propagation (note that periodic boundary conditions are used). The
presence of such coherent structures is related to the values of $\alpha$
and $\gamma$ and not to the particular random phases selected. 
%%%%%%%%%%%%%%%%%%%
%%%%%%%%%%%%%%%%%%%
%%%%%%%%%%%%%%%%%%%
\begin{figure} 
\epsfxsize=8.5cm 
\epsfbox{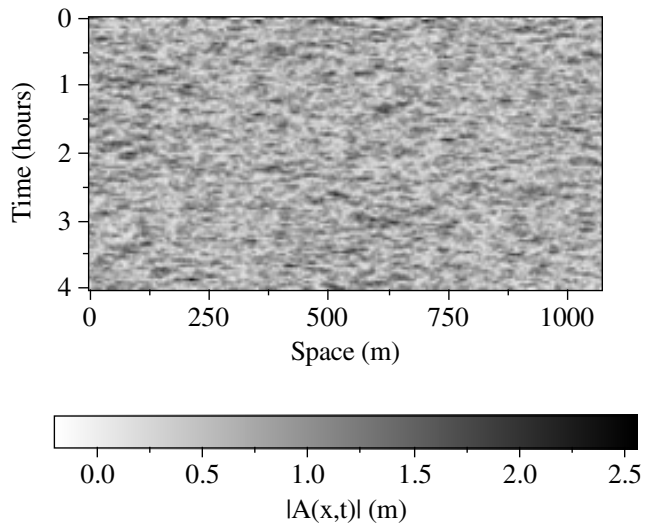}
\caption{$|A(x,t)|$ from numerical simulation of the NLSE. The initial condition
is characterized by a Lorentian spectrum with $\gamma=1$, $\alpha=0.02$.
The field in the $x-t$ plane appear to be random with out any evidence
of coherent structure.} \label{fig figura2}
\end{figure}
%%%%%%%%%%%%%%%%%%%
Every random
realization shows similar results even though the resulting coherent
structures may have different velocity and amplitude. The nonlinear stage of
MI is therefore responsible for the formation of such coherent structures in
the $x-t$ plane. Indeed it is possible to show that the NLSE has periodic
solutions such as, for example, breathers or unstable modes \cite
{Akhm,OSBPLA}. These solutions, which are linearly unstable, are
nevertheless very robust. Moreover they can grow up to more than three times
the initial unperturbed solution and therefore have also been addressed as
simple models for freak waves \cite{OSBPLA,DT}. In contrast to solitons that
have constant amplitude in time, these unstable modes are characterized by a
continuous exchange of energy among the Fourier modes. The energy is
transferred from one mode to another and back again: the process is
completely reversible and therefore coherent structures persist in physical
space. 
We stress that these kinds of solutions appear naturally from initial
conditions with random phases.
We now consider a case with $\gamma=1$ and $\alpha=0.02$ ($p/k_0\simeq 0.58$
and $H_s k_0/2 \simeq 0.12$); note that decreasing $\gamma$ has the effect
of making the spectrum broader. This case, according to the stability
criteria (see  Fig. \ref{fig inst1}), should not present any instability
because of the effect of Landau damping. We show the output of our numerical
simulation in Fig \ref{fig figura2}. In the $x-t$ plane, as a confirmation
of what we have previously stated, there is no evidence of any coherent
structure; numerical simulations with different random phases are in
accordance with the result just shown.

In conclusion, the numerical simulations show that the presence of the coherent
structures in the $x-t$ plane is related to the instability of the wave field.
The Landau damping phenomenon suppress the instability and prevent the
formation of coherent structures. Many physical questions remain open.
For example it would be interesting to
investigate the case of a two dimensional wave field. It is well known that
the NLSE in 2+1 is not integrable and the dynamics of coherent structures is
still far from being understood. Numerical simulations with the fully
nonlinear Euler equations are also under consideration in order to extend
the validity of the result. Concerning the problem of wave forecasting, we
may state that if one is interested in predicting freak waves arising from
the MI, this new form of the kinetic equation, which includes nonhomogeneous
effects should be considered. %
%%%%%%%%%%%%%%%%%%%%%%%%%%%%%%%%%%%%%%%%%%%%%%%%%%%%%%%%%%%%%%

P. Janssen and D. Grasso are acknowledged for valuable discussions. M. O. was supported by
a Research Contract from the Universit\`{a} di Torino. This work was
supported by the O.N.R. (T. F. Swean, Jr.) and by the Mobile Offshore Base
Program of O.N.R. (G. Remmers). Consortium funds and Torino University funds
(60 \%) are also acknowledged.

%%%%%%%%%%%%%%%%%%%%%%%%%%%%%%%%%%%%%%%%%%%%%%%%%%%%%%
%
\end{multicols} 
\end{document}